\documentclass[aps,pra,amsmath,superscriptaddress,reprint,floatfix,notitlepage,balancelastpage,twocolumn,reprint]{revtex4}
\usepackage[colorlinks,urlcolor=red,citecolor=red]{hyperref}
\usepackage{graphicx,subfigure} 
\usepackage{epstopdf}
\usepackage{amsthm} 

\usepackage{tikz}

\let\baraccent=\= 
\renewcommand{\=}[1]{\stackrel{#1}{=}} 

\theoremstyle{definition}

\theoremstyle{remark}

\newcommand{\as}{a_s}

\newcommand{\tc}{{T_{\rm c}}}

\newcommand{\phdag}{{\phantom{\dagger}}}

\newcommand{\kf}{k_{\rm F}}

\newcommand{\kb}{k_{\rm B}}

\newcommand{\bk}{{\bf k}}
\newcommand{\bq}{{\bf q}}
\newcommand{\bp}{{\bf p}}

\newcommand{\ph}{\hat{p}}

\newcommand{\br}{{\bf r}}

\newcommand{\be}{\begin{equation}}
\newcommand{\ee}{\end{equation}}
\newcommand{\bea}{\begin{eqnarray}}
\newcommand{\eea}{\end{eqnarray}}
\newcommand{\bse}{\begin{subequations}}
\newcommand{\ese}{\end{subequations}}

\begin{document}

\title{Renormalization group approach to the normal phase of 2D Fermi gases}
\author{S. Laalitya Uppalapati}
\email{suppal5@lsu.edu}
\author{Daniel E. Sheehy}
\email{sheehy@lsu.edu}
\affiliation{Department of Physics and Astronomy, Louisiana State University, Baton Rouge, LA, 70803, USA}
\date{\today}
\begin{abstract} 
We present results on the effect of
short-range, attractive interactions on the properties of balanced 2D
Fermi gases in the non-superfluid (normal) phase. Our approach
combines the renormalization group (RG) with perturbation theory,
yielding observables such as the equation of state and
compressibility. We find good agreement with recent experiments that
measured the equation of state in trapped gases in the balanced regime, showing that these results are consistent with logarithmic
corrections in the equation of state.

\end{abstract}


\maketitle

\section{Introduction}
Correlated solid-state materials are challenging to describe theoretically, 
with a full understanding
requiring an accounting of the interplay between electronic degrees of freedom, complex crystal 
structures, phonons and disorder. 
 Cold atomic gases have emerged as a simpler analogue of
 solid state systems, exhibiting tunable
dimensionality, interaction stength and spin
polarization~\cite{Bloch08,Giorgini08}. 
Here we focus on atomic fermions 
confined to two dimensions (2D), a subject of much recent
experimental and theoretical interest, with Ref.~\cite{LevinsenReview15} providing a review.
Some experimental highlights include early work that detected the 2D regime for atomic fermions~\cite{Modugno03},
the measurement of pairing correlations~\cite{Feld11,Sommer12} and polaronic correlations~\cite{Zhang12},
and the study of collective modes~\cite{Vogt12} of interacting 2D (and quasi-2D) Fermi gases.
Subsequent experiments observed the crossover between fermionic and bosonic correlations for
quasi-2D gases~\cite{Makhalov2014}, evidence of pair  condensation~\cite{Ries15} and
the Beresinzkii-Kosterlitz-Thouless (BKT)~\cite{Berezinskii72,Kosterlitz73} superfluid
transition~\cite{Murthy15}. 
 Experiments also observed  beyond mean-field effects in radio-frequency spectra of 2D gases~\cite{Cheng2016}, and
probed the imbalanced regime, in which a global magnetization or population imbalance is imposed~\cite{Mitra2016}.
More recent experiments have explored a quantum anomaly in collective modes of 2D trapped gases~\cite{Holten2018,Peppler2018},
and the development of uniform box potentials for cold atoms~\cite{Gaunt,Mukherjee17} has led to the study of quasi-2D uniform
gases~\cite{Hueck18}.

\begin{figure}[ht!]
     \begin{center}
 \includegraphics[width=85mm]{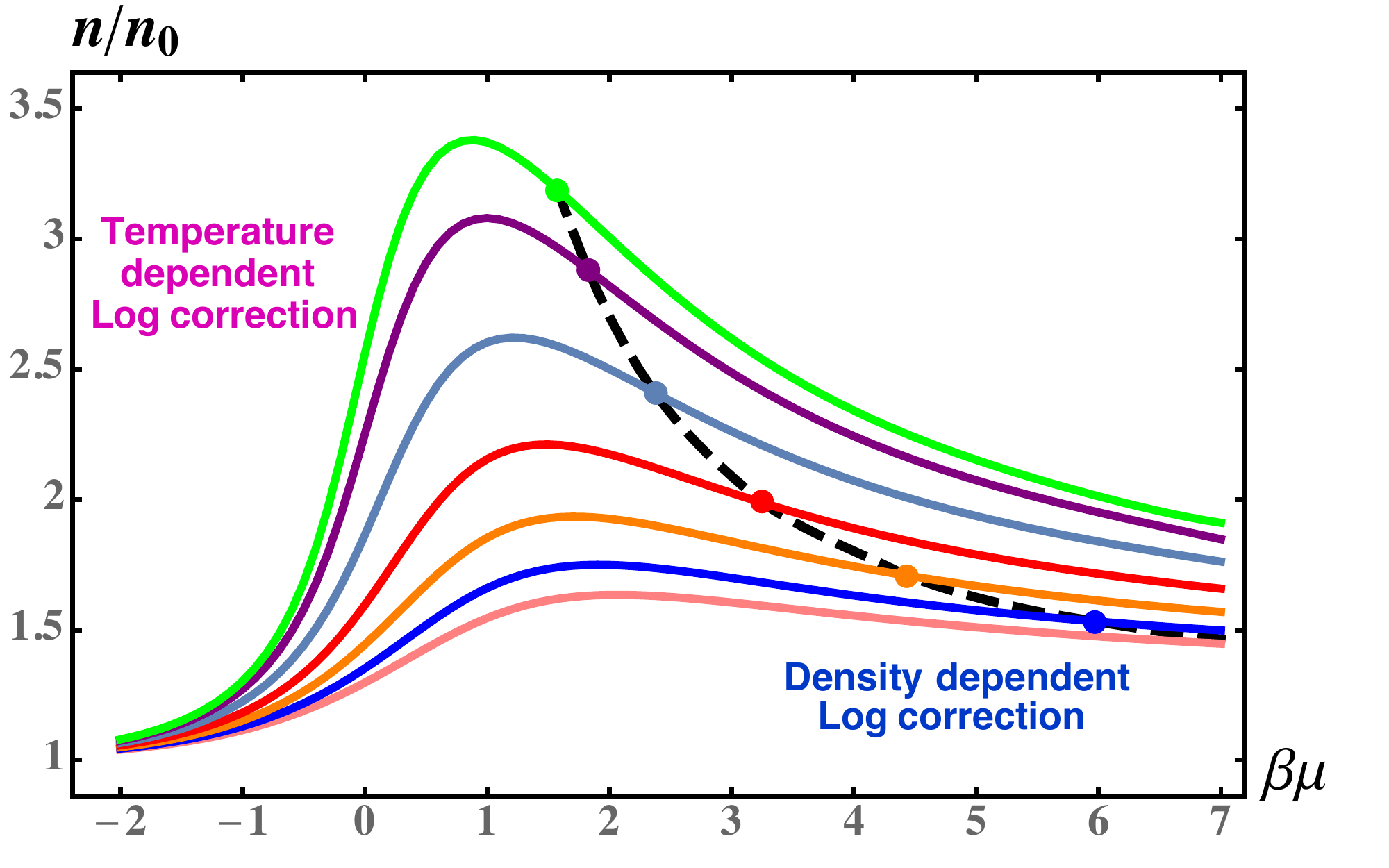}
%
    \end{center}\vspace{-.5cm}
    \caption{
        (Color Online) Density 
(normalized to the noninteracting density $n_0$) of a 2D Fermi gas 
vs. chemical potential (normalized to $\kb T = \beta^{-1}$), for
dimensionless 
coupling values $\beta E_b$ that,
from top to bottom, are $0.84, 0.76, 0.62,0.47,0.35,0.26,0.20$, showing a peak
in the normalized density for intermediate values of $\beta \mu$ that is reminiscent of
recent experiments~\cite{Fenech16,Boettcher16}.
These curves are computed using Eq.~(\ref{eq:finalmuvsn}) with the effective coupling approximately
interpolating between $\lambda_{\rm eff}~\sim 1/\ln T$ in the left part of
the figure and $\lambda_{\rm eff}~\sim 1/\ln n$ in the right part of the figure, as
argued in Sec.~\ref{eq:SEC:rg}.
For comparison, the dashed line shows the mean-field 
transition temperature (given in Eq.~(\ref{eq:mftc}) below) to the superfluid phase, suggesting a connection between the peak in $n/n_0$ 
and the onset of pairing correlations. 
     }
\vspace{-.25cm}
   \label{fig:one}
\end{figure}

Theoretically, the study of interacting 2D Fermi gases has been active for decades, motivated by the application to condensed
matter systems.  Randeria and collaborators provided important early work on pairing and superfluidity in 2D Fermi gases~\cite{Randeria89,Randeria90}, with
Petrov and collaborators addressing this system in the cold-atom context~\cite{Petrov03}.  Further work on the BEC-BCS crossover of the superfluid
state has investigated the importance of transverse levels of the confining potential of these quasi-2D gases~\cite{Fischer13}.  

In this paper our main interest is in recent experiments~\cite{Fenech16,Boettcher16}
 in the {\em nonsuperfluid\/} phase of 2D Fermi gases that reveal
the importance of interactions in the measured density equation of state, despite the lack of any long-range order.   
Both groups observed a bump in the density $n$ (normalized to the noninteracting density $n_0$) as a function of
the chemical potential $\mu$ (normalized to $\beta^{-1} = \kb T$, with  $\kb$  the Boltzmann's constant and $T$ the temperature).  
Our theoretical results (depicted in Fig.~\ref{fig:one}), are based on a renormalization group (RG) approach.  This approach
shows that short-ranged interactions are marginal in this system, implying logarithmic corrections relative to the noninteracting
case (as also found in other systems with marginal interactions, like graphene~\cite{Sheehy2007}).  We find that the
bump can be traced to
a crossover between a density-dependent logarithmic correction to the noninteracting equation of state
at large $\beta \mu$ to a temperature-dependent logarithmic correction at small $\beta \mu$.

This paper is organized as follows.  In Sec.~\ref{sec:hamiltonian}, we introduce a Hamiltonian
for fermions in two spatial dimensions with short-range attractive interactions.
In Sec.~\ref{eq:SEC:rg}, we introduce our RG approach, which involves combining a renormalization
group equation for the effective coupling with scaling equations for observables as well
as a choice for when to terminate the RG flow.
In Sec.~\ref{sec:IRG}, we improve our RG calculation by numerically integrating the RG
equation.  We then compare our results for the density vs. chemical potential to
the experimental results of Refs.~\cite{Fenech16,Boettcher16}.  
In Sec.~\ref{OTQ}, we extend our results to other thermodynamic observables such as the
system pressure and compressibility, and in 
 Sec.~\ref{sec:CR} we provide some concluding remarks.  

\section{Hamiltonian}
\label{sec:hamiltonian}
Our starting point is the following Hamiltonian for a balanced gas of 2D fermions of mass $m$ with
short-ranged, attractive interactions with coupling constant $\lambda$:
\label{Eq:hamiltonian}
\bea
&& H =\int d^2 r \, \psi_\sigma^\dagger(\br) \big[ \frac{\ph^2}{2m} -\mu \big] \psi_\sigma^\phdag(\br)
\\
&& + \lambda \int d^2 r \, \psi_\uparrow^\dagger(\br) \psi_\downarrow^\dagger(\br) 
\psi_\downarrow^\phdag(\br) \psi_\uparrow^\phdag(\br), \nonumber
\eea
where $\ph=-i\nabla$ is the momentum 
operator.   (Here and below we take Planck's constant $\hbar = 1$, and 
below we also take the Boltzmann constant $\kb = 1$.).
The field operators $\psi_\sigma(\br)$ annhilate a fermion of spin $\sigma=(\uparrow,\downarrow)$ at spatial position $\br$ and 
satisfy the anticommutation relations $\{\psi_\sigma(\br),\psi_{\sigma'}^\dagger(\br')\} = \delta_{\sigma,\sigma'}\delta^{(2)}(\br-\br')$.  This Hamiltonian must be regularized in
the ultraviolet (UV) to avoid divergences, which in the Fourier series representation implies that momenta are
restricted to $|\bp|<\Lambda$.  The cutoff scale $\Lambda$ and interaction parameter
are related to the two-body binding energy $\varepsilon_b = \frac{1}{ma_s^2}$ (with $\as$ the two-body scattering
length) via: 
\be
\label{Eq:renormalization}
\frac{1}{\lambda}=-  \sum_\bk \frac{1}{2\epsilon_k + \varepsilon_b},
\ee 
where $\epsilon_k=\frac{k^2}{2m}$ is the kinetic energy.

\section{Renormalization Group} 
\label{eq:SEC:rg}
Our aim is to investigate how interactions impact observables of a 2D interacting Fermi gas in the nonsuperfluid
phase, for $T>\tc$.  To do this, we use the Renormalization Group (RG) framework which allows us to relate
efficiently determine the effect of high-momentum degrees of freedom on the low-energy theory.
We first introduce the finite-$T$ action $S = S_0 + S_1$ corresponding to the functional integral
representation of the partition function associated with the Hamiltonian $H$.  Here, $S_0$ is the
noninteracting part of the action: 
\bea
S_0=\frac{1}{\beta} \int_{\bk<\Lambda} \frac{d^2k}{(2\pi)^2} 
\sum_{\bk,\omega} \psi_{\sigma}^\dagger (\bk,\omega) \big[-i\omega+\xi_\bk \big] \psi_\sigma (\bk,\omega),
\eea
where we converted the momentum sum to an integration, with the system area $A$ set to unity.  Here,
 $\psi_\sigma(\bk,\omega)$ is the Fourier-transformed Grassmann field with 
 $\bk$ momentum 
and  $\omega = \frac{\pi}{\beta}(2n+1)$ is a fermionic Matsubara frequency and 
 $\xi_\bk=\frac{k^2}{2m}-\mu$.  The  interaction part of the action, $S_1$, is:
\be
S_1 = \sum_{P,K,Q}\lambda 
\psi_\uparrow^\dagger(P_+) \psi_\downarrow^\dagger(-P_-) \psi_\downarrow(K_+) \psi_\uparrow(-K_-) ,
\label{eq:sonee}
\ee
with $P=(\bp,i\omega)$ a three-vector including Matsubara frequency, $\sum_P = T\sum_\omega  \int_{\bp<\Lambda}\frac{d^2 p}{(2\pi)^2}$, 
and $P_{\pm} = P\pm \frac{1}{2}Q$ (and similarly for $K_\pm$).
 
The RG approach involves performing a partial trace over high-momentum degrees of freedom.  
To implement the RG, we  define 
a new cutoff 
\be
\Lambda' = \Lambda/b,
\ee 
with renormalization constant $b$ such that $b>1$,
write the fermion fields as 
\be
\label{Eq:splitpsi}
\psi_\sigma(\bk,\omega) = \psi^<_\sigma(\bk,\omega) \Theta(\Lambda'-k) +  \psi^>_\sigma(\bk,\omega)  \Theta(k-\Lambda'),
\ee
and then trace (perturbatively in $\lambda$) over the high-momentum fields $\psi^>_\sigma(\bk,\omega)$ with 
  wavevectors between $\Lambda'$ and $\Lambda$, yielding an action for the low-momentum
fields $\psi_\sigma^<(\bk,\omega)$. 
 We then 
define new wavevectors $\bk'$ and new frequencies $\omega'$  via:
\bse
\label{Eq:rescalingsub}
\bea
\bk = b^{-1} \bk',
\\
\omega = Z_T \omega',
\eea
\ese
and define renormalized fermion fields $\psi_\sigma$ via 
\be
\label{Eq:psirescaling}
\psi_\sigma^<(b^{-1} \bk',Z_T \omega') = Z_\psi \psi_\sigma(\bk',\omega').
\ee
with $Z_T$  and $Z_\psi$ determined by demanding that $S_0$ have its original form.
  Indeed, it is easy to see that the noninteracting action is invariant under this RG tranformation if we take $Z_T = b^{-2}$ and $Z_\psi = b^3$,
and renormalize other energy scales analogously to
$\omega$, with $\mu =  Z_T \mu'$ and $T = Z_T T'$.

 To zeroth order in the coupling $\lambda$, 
which is the tree-level RG, $S_1$ is also invariant under the RG transformation, reflecting the marginal nature of the interactions.  
 To
quadratic order in the coupling, however, we find that the action is modified, with  the  change in the effective coupling for
the low-momentum fields given by 
\be
\label{Eq:deltalambda}
\delta \lambda \equiv  - \lambda^2 T \sum_\Omega \int_> \frac{d^2 q}{(2\pi)^2} G(\bq,\Omega) G(-\bq,-\Omega) ,
\ee
with $G(\bq,\Omega) = \frac{1}{i\Omega -(\epsilon_q -\mu)}$ the noninteracting Green's function and the $\bq$ integration over
momenta $\Lambda/b<|\bq|<\Lambda$.  In the simplest approximation, in which we set external momenta and frequencies to zero and let $T \rightarrow 0$, $\mu \simeq 0$ (assumptions that we shall relax below), this results in the following RG equation: 
\be
\label{RGatZeroTemp}
\frac{d\lambda(b)}{d\ln b}=-\frac{m}{2\pi}\lambda(b)^2.
\ee
Since the bare coupling constant $\lambda <0$, for attractive interactions,
this equation describes the progressive increase in the magnitude of the interactions under the RG transformation, as seen by the solution 
\be
\label{eq:lambaRGresult}
\lambda(b) = \frac{\lambda}{1+ \frac{m}{2\pi } \lambda \ln b}.
\ee
showing  a slow growth of $|\lambda(b)|$ with increasing $b$, as the RG procedure is iterated.  Although $\lambda(b)$ given in Eq.~(\ref{eq:lambaRGresult})
{\em diverges\/} for sufficiently large $b$, in practice we will terminate the RG flow before this occurs.

The next step is to derive scaling relations for physical observables.  Of particular interest is the connection between the atom density $n$ and 
the system chemical potential $\mu$, measured in experiments on trapped 2D gases
when the local density approximation (LDA) is invoked.  Using the definition $n = \sum_\sigma\psi^\dagger(\br)\psi(\br)$ for the density, it is straightforward
to find that the renormalized density satisfies $n(b) = b^2 n$, which is expected since the RG transformation corresponds to a coarse graining in real space.  
The preceding considerations imply the following relation:
\be
\label{Eq:murelation}
\mu(T,n,\lambda) = b^{-2} \mu\big(T(b),n(b), \lambda(b)\big),
\ee
between the physical chemical potential (on the left side, as a function of the physical temperature, density and coupling) and the chemical potential in the 
renormalized system (on the right side).  A more compact version of this equation is $\mu =  b^{-2}\mu_R$.
 We proceed by combining a perturbative calculation of the system chemical
potential in the renormalized system with a choice for the renormalization parameter $b$ (the renormalization condition at which we terminate the RG flow).
  The justification for
this is that, since $T(b)$ and $n(b)$ grow quadratically under the RG transformation, the renormalized system is in a high
temperature and high density regime, justifying a perturbative analysis based on the Hartree-Fock approximation, with the result:
\be
\mu(T,n,\lambda) = T \ln \Big[ {\rm e}^{\frac{\pi  n}{mT}} -1\Big] + \frac{1}{2}\lambda n.
\ee
We thus use this formula for the right side of Eq.~(\ref{Eq:murelation}), by plugging in $T(b)$, $n(b)$, and $\lambda(b)$.
Our renormalization condition, or choice for $b$, is dictated by the fact that our approach breaks down when $T(b)$ or 
$n(b)$ become too large, reaching the  UV scale
(determined by $\Lambda$) such that 
the approximations leading to Eq.~(\ref{RGatZeroTemp}) no longer hold. 
Since $T(b)$ and $n(b)$ scale identically under the RG transformation, the dimensionless ratio 
\be
\label{eq:nbar}
\bar{n} \equiv \frac{\pi n}{mT}
\ee
 is
independent of the RG 
transformation, and we have different behavior depending on whether $\bar{n}>1$ or $\bar{n}<1$. 

 We start with the regime of $\bar{n}\gg 1$, which we call the density-dominated regime.   In this regime, we take the renormalization
condition $n(b^*) = \Lambda^2/(2\pi)$ by when the renormalized density reaches
the density scale controlled by the cutoff (note that, since we took $\hbar\to 1$, the cutoff
$\Lambda$ has units of inverse length).  
  This yields an effective coupling in the renormalized system  given by 
  $\lambda_{\rm eff} \equiv \lambda(b^*) = - \frac{2\pi}{m} \frac{1}{\ln \sqrt{2\pi n \as^2}}$.   With this renormalization condition,
  our result for the chemical potential is:
\bea
\label{eq:muvsn}
\mu &=& T \ln \Big[ {\rm e}^{\frac{\pi n}{mT}} -1\Big] + \frac{1}{2}\lambda_{\rm eff} n,
\\
&=& T \ln \Big[ {\rm e}^{\frac{\pi  n}{mT}} -1\Big] - \frac{\pi n}{m} \frac{1}{\ln \sqrt{2\pi n \as^2}} ,
\label{eq:muvsn2}
\eea
where in the second line we plugged in the our result for the effective renormalized coupling that holds in the regime $\bar{n} \gg 1$.
Thus, we find that $\mu$ in the high-density regime is given, approximately, by its noninteracting value 
(the first term on the right side of Eq.~(\ref{eq:muvsn2})) with a correction going as $\sim 1/\ln n$, consistent
with the results of Bloom~\cite{Bloom74}, who analyzed the zero-temperature 2D Fermi gas within a diagrammatic approach.

\begin{figure}[ht!]
     \begin{center}
 \includegraphics[width=85mm]{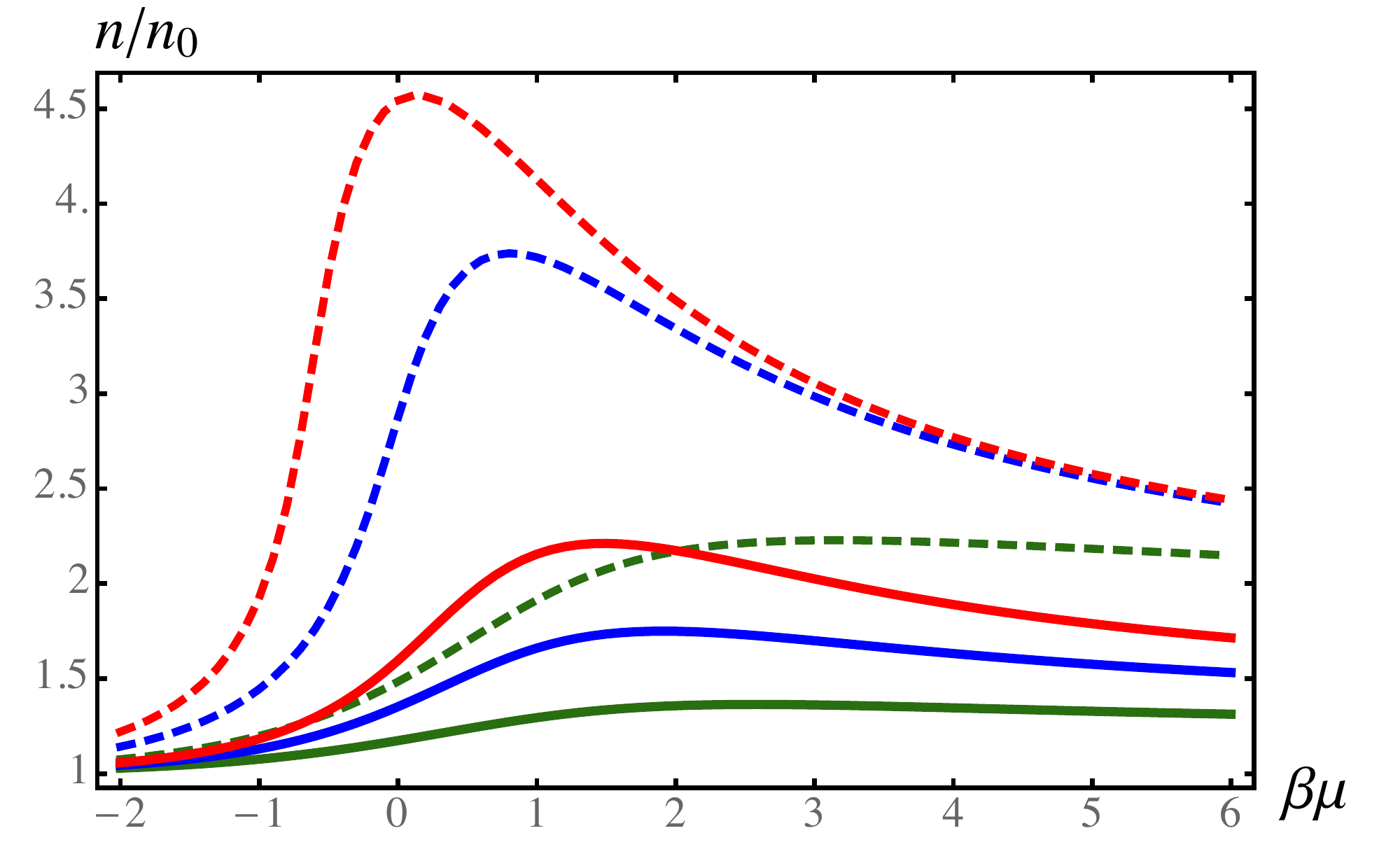}
%
    \end{center}\vspace{-.3cm}
    \caption{
        (Color Online) Density (normalized to the noninteracting density) vs. chemical potential (normalized to  $\kb T = \beta^{-1}$)
 within the simplified RG scheme in Eq.~(\ref{eqnaiveRGresult}) (dashed)  and
via integration of the RG equation in Eq.~(\ref{eq:finalmuvsn}) (solid).  The coupling values for each are $\beta E_b = 0.47$ (top, red)
$\beta E_b = 0.26$ (middle, blue), and $\beta E_b = 0.06$ (bottom, green).  While the two sets of curves agree qualitatively, exhibiting a ``bump'' 
as a function of normalized chemical potential, they are quantitatively different.  
     }
\vspace{-.25cm}
   \label{fig:two}
\end{figure}

In the preceding, the fact that the logarithmic term in Eq.~(\ref{eq:muvsn2}) depended on density was traced to the fact
that we assumed $\bar{n}\gg 1$, so that $n(b)$ reaches the UV scale first, defining the renormalization condition.  
In the opposite regime of $\bar{n}\ll 1$, the temperature-dominated regime, we instead choose 
$T(b^*)= \frac{\Lambda^2}{2m}$, defined by when the renormalized temperature reaches the typical kinetic energy scale determined by
the cutoff.  This gives, for the renormalized coupling, $\lambda_{\rm eff} = - \frac{2\pi}{m} \frac{1}{\ln \sqrt{2m T \as^2}}$. 
   Once again, we combine this with the perturbative result for the chemical
potential, so that Eq.~(\ref{eq:muvsn}) holds but with the interaction correction $\sim 1/\ln T$:
\be
\mu  \simeq T \ln \Big[ {\rm e}^{\frac{\pi  n}{mT}} -1\Big]  - \frac{\pi n}{m} \frac{1}{\ln \sqrt{2m T \as^2}},
\ee
with the only difference being the argument of the logarithmic correction.  
 These RG results suggest that the bump in the normalized density vs. chemical potential observed in ~\cite{Fenech16,Boettcher16}
 comprises a crossover between the density-dominated and temperature-dominated regimes, resulting in a crossover
 between
 a density dependent logarithmic correction at $\bar{n}\gg 1$ to a temperature-dependent
logarithmic correction at $\bar{n}\ll 1$.  To illustrate this, in the dashed lines of Fig.~\ref{fig:two} we plot the density
$n$ normalized to the noninteracting density $n_0$, obtained by numerically solving Eq.~(\ref{eq:muvsn}) with effective coupling
 $\lambda_{\rm eff} = -\frac{2\pi}{m}\frac{1}{\ln \sqrt{2m Ta^2f(\bar{n})}}$ 
where $f(\bar{n})  = \sqrt{\bar{n}^2+1}$ an approximate function that 
interpolates between these two limiting cases.  This leads to (recall $\bar{n}$ is defined in Eq.~(\ref{eq:nbar})):
\be
\label{eqnaiveRGresult} 
\beta \mu = \ln \big[ {\rm e}^{\bar{n}} - 1\big] - \frac{\bar{n}}{ \ln \sqrt{2m Ta^2f(\bar{n})}} ,
\ee
which we plot as dashed lines in Fig.~\ref{fig:two} for various values of the coupling strength.  Remarkably, this simple 
formula indeed qualitatively reproduces the ``bump'' behavior seen in experiments.

 Since our chosen interpolation function appears in the argument of the logarithm in Eq.~(\ref{eqnaiveRGresult}), 
its precise form is expected to be relatively unimportant as long as it correctly matches to these limits.   
Nonetheless, our interpolation between these limits is ad-hoc, and in the next section we seek a more rigorous
way to compute interaction effects in 2D Fermi gases using the RG approach.

\section{Integrating the RG equation}
\label{sec:IRG}
Our scheme to terminate the RG flow is only valid in the regimes of $\bar{n}\gg 1$ or $\bar{n}\ll 1$ where there is a 
well defined procedure to define the renormalization condition.  Although our ad-hoc interpolation between these 
regimes is qualitatively valid, correctly yielding the bump in the equation of state, it is not accurate enough to
attempt a direct comparison
with experimental data.   Our next task is to improve our scheme by integrating the full RG equation with the aim of comparing to experiments.   

To do this, we need to relax the assumptions leading to the approximate RG relation Eq.~(\ref{RGatZeroTemp}).
We begin by generalizing our effective action to allow for a frequency and momentum dependence to the coupling $\lambda$, 
which amounts to replacing $\lambda\to \lambda(q,\Omega) = \lambda(Q)$ in Eq.~(\ref{eq:sonee}):
\be
\label{eq:soneeNEW}
S_1 = \sum_{P,K,Q}\lambda(Q)
\psi_\uparrow^\dagger(P_+) \psi_\downarrow^\dagger(-P_-) \psi_\downarrow(K_+) \psi_\uparrow(-K_-) .
\ee
The reason for doing this is as follows: 
Although the bare coupling is $Q$-independent, we shall see that such a dependence is generated by the RG flow.  Here, we have assumed
that the coupling only depends on the magnitude (not the direction) of the momentum ($q$).  

To derive the RG equation, we start with the tree-level case.  In this case, 
the rescaling steps in Eq.~(\ref{Eq:rescalingsub}) imply that the new coupling is
\bea
\label{newcouplingtree}
\lambda_{\rm new}(q,\Omega) &=& \lambda(b^{-1} q,b^{-2}\Omega) 
\\
&=& 
 \lambda(q,\Omega) - q \frac{\partial\lambda}{\partial q} d\ln b -2\Omega \frac{\partial \lambda}{\partial \Omega} d\ln b.
\eea
with the second line applying in the case of an infinitesimal RG transformation, obtained by taking $b^{-1} \simeq 1-d\ln b$ and Taylor expanding to leading order.  
Next, we include the quadratic order perturbative RG correction, that is analogous to Eq.~(\ref{Eq:deltalambda}) but with the momentum and frequency-dependent coupling.
This leads to a differential equation for the running effective coupling:
\bea \label{eq:RGPW}
&&\hspace{-.2cm}\frac{d\lambda(q ,\Omega , b)}{d\ln b}  
=  - q \frac{\partial\lambda}{\partial q} -2\Omega \frac{\partial \lambda}{\partial \Omega} 
- f(q,\Omega) \lambda^2, \\
&&\hspace{-.2cm}f(q,\Omega) \equiv   g \epsilon_\Lambda \int \frac{d\theta}{2\pi} \frac{1}{2(\epsilon_\Lambda - \mu_R + \frac{q^2}{8m})- 
i\Omega}
\label{eq:functionf}
\\
&&\times 
 \sum_{\alpha = \pm 1} \tanh \Big[\frac{\epsilon_\Lambda - \mu_R + \frac{q^2}{8m} + \alpha \frac{1}{\sqrt{2m}}q\sqrt{\epsilon_\Lambda}\cos\theta}
{2T_R}\Big].\nonumber 
\eea
where $g =\frac{m}{2\pi} $ is the density of states, $\epsilon_\Lambda = \Lambda^2/2m$ is a cutoff-dependent energy scale, and $\mu_R=\mu b^2$ and $T_R = Tb^2$ are the running
(renormalized) chemical potential and temperature.  Note that on the right side 
of Eq.~(\ref{eq:RGPW})
we suppressed the arguments of the coupling, which are the same as 
on the left side.  To solve this equation for the scale-dependent coupling $\lambda(q,\Omega,b)$,
 we note that, at tree-level, the RG procedure amounts to a rescaling of momenta and frequencies, as
seen in Eq.~(\ref{newcouplingtree}).  This motivates the ansatz $\lambda(q,\Omega,b) = \lambda_0(b^{-1}q,b^{-2}\Omega,b)$.  Plugging this into
Eq.~(\ref{eq:RGPW}), we find that 
$\lambda_0(q,\Omega,b)$ satisfies the simpler differential equation 
\be
\label{eq:lambdaN}
\frac{\partial \lambda_0(q,\Omega,b)}{\partial \ln b} = - f(bq,b^2\Omega) \big(\lambda_0(q,\Omega,b)\big)^2,
\ee
which we need to solve, subject to the initial condition $\lambda_0(q,\Omega,1) = \lambda$ (the momentum independent bare coupling).
The latter satisfies: 
\be
\label{Eq:renormalizationN}
\frac{1}{\lambda} = -g\int_0^{\epsilon_\Lambda} dx \frac{1}{2x+\varepsilon_b} ,
\ee
which is equivalent to Eq.~(\ref{Eq:renormalization}).
  The solution to Eq.~(\ref{eq:lambdaN})  is:
\be
\frac{1}{\lambda_0(q ,\Omega , b)} - \frac{1}{\lambda} = \int_1^{b}\frac{db_1}{b_1}  f(b_1 q,b_1^2\Omega).
\ee
Upon changing variables in the integration to $x = \epsilon_\Lambda/b_1^2$, plugging in Eq.~(\ref{Eq:renormalizationN})  and approximating $b\to \infty$ in the
final expression, we obtain (after also setting the frequency $\Omega \to 0$):
\bea
\label{eq:lambdanought}
&&\frac{1}{\lambda_0(q ,0)} 
\\
&&
=\int_0^{2\pi} \frac{d\theta}{2\pi} \int_0^\infty d\epsilon \Big[ \frac{1-\eta_f(\chi_+)-\eta_f(\chi_-)}{\chi_+ +\chi_-}-\frac{1}{2\epsilon+\epsilon_b}\Big]
\nonumber
\eea
where $\chi_\pm=\epsilon+1/8 q^2-\mu \pm q \sqrt{\frac{\epsilon}{2}} \text{cos} \theta$ and $\eta_f (x)=1/(1+e^{x/T})$ is the Fermi distribution function. 
Here, we defined $\lambda_0(q ,0) \equiv \lim_{b\to \infty} \lambda_0(q ,0,b)$.

 The next step is to compute the single-particle Green's function $G_\sigma(\bk,\omega)$
for our system (for spin $\sigma$), given by the average 
$-\langle \psi_\sigma(\bk,\omega)\psi_\sigma^\dagger(\bk',\omega')\rangle 
= \delta^{(2)}(\bk-\bk') \beta \delta_{\omega,\omega'} G_\sigma(\bk,\omega)$.  To do this, we use 
the scaling relation (following from Eq.~(\ref{Eq:psirescaling})):
\be
\label{greenscaling}
G_\sigma(\bk,\omega) =  b^2G_{\sigma,R}(b\bk,b^2\omega),
\ee
where the Green's function of the renormalized system on the right side of this formula should be computed using the momentum and frequency dependent effective coupling $\lambda(q,\Omega,b) = \lambda_0(b^{-1}q ,b^{-2}\Omega , b)$ with
$\lambda_0(q,\Omega,b)$ given (at large $b$) in Eq.~(\ref{eq:lambdanought}).  Within standard perturbation theory, the Green's function of the renormalized system 
satisfies 
\be
\label{eq:greenself}
G_{\sigma,R}(\bk,\omega) = \frac{1}{i\omega - \xi_\bk - \Sigma_\sigma(\bk,\omega)},
\ee
with the self energy 
\bea
\label{Eq:selfenergyPre}
\Sigma_\sigma(\bk,\omega)\!\! &=&\!\! 
T\sum_{\omega'} \int \frac{d^2 k'}{(2\pi)^2} \lambda(|\bk+\bk'|,\omega+\omega',b)G_{\bar{\sigma}R}(\bk',\omega'),
\nonumber
\\
&\simeq& \lambda(k,0,b) n_{R,\bar{\sigma}},
\eea
where $\bar{\sigma}$ is the opposite spin to $\sigma$.   In the second line we neglected the frequency dependence of the renormalized coupling and approximated 
$\lambda(|\bk+\bk'|,0,b) \simeq \lambda(k,0,b) $, which holds if the coupling is only weakly momentum dependent.  
Within these approximations, the frequency and momentum sum
in Eq.~(\ref{Eq:selfenergyPre})  simply give the density of spin-$\bar{\sigma}$ fermions (equal to one-half the total 
fermion density).  Now, combining this with Eqs.~(\ref{eq:greenself}) and 
Eq.~(\ref{greenscaling}), 
we finally arrive at:
\be
\label{eq:finalgreen}
G_\sigma(\bk,\omega) = \frac{1}{i\omega - \xi_\bk - \frac{1}{2}n \lambda_0(k,0)},
\ee
 This single-particle Green's function describes interacting
fermions with an effective momentum-dependent chemical potential given by $\mu - \frac{1}{2}n\lambda_0(k,0)$ with
the Fermi surface located at momentum $\kf$ satisfying
\be
0=\frac{\kf^2}{2m} -\mu + \frac{1}{2} n\lambda_0(\kf,0).
\ee
According to Luttinger's theorem~\cite{Luttinger}, the fermion density at low temperatures is governed by the volume in 
momentum space of this Fermi surface via $n = \frac{\kf^2}{2\pi}$ (as in the noninteracting case).  We therefore argue
that it is approximately valid  to replace $\lambda_0(k,0) \to \lambda_0(\kf,0)$ in Eq.~(\ref{eq:finalgreen}) for
the purposes of calculating the total density, which leads to:
\be
n = \frac{mT}{\pi} \ln \Big[ {\rm e}^{(\mu+ \frac{1}{2}n \lambda_{\rm eff})/T} +1\Big] ,
\ee
or, equivalently, 
\be
\label{eq:finalmuvsn}
\beta \mu  = \ln \Big[ {\rm e}^{\pi n/(mT)} -1\Big] - \frac{1}{2} n\lambda_{\rm eff},
\ee
 where we defined the effective coupling at the Fermi level $\lambda_{\rm eff} \equiv \lambda_0(\kf,0)$.  

To obtain the density vs. chemical potential within this theory, we numerically
evaluate the wavevector dependent coupling Eq.~(\ref{eq:lambdanought})
(that also depends on the fermion density via $\kf$) and then numerically solve
Eq.~(\ref{eq:finalmuvsn}).  In Fig.~\ref{fig:one} we plot the resulting 
density, normalized to the noninteracting density, as a function of
$\beta \mu$ for various interaction strengths.  

In Fig.~\ref{fig:two},
we compare these results (shown as solid curves) with the results of our
ad-hoc interpolation scheme from Sec.~\ref{eq:SEC:rg}  (shown as dashed curves). We argue
that the qualitative agreement between the two sets of curves implies the overall
validity of the basic picture of the bump as a crossover between temperature-dependent
log correction in the temperature-dominated regime and a density-dependent log correction in 
the density-dominated regime.  

At sufficiently low temperatures (at fixed density), or at sufficiently high density (at fixed temperature)
 a 2D attractive Fermi gas is expected to enter a paired superfluid (i.e., superconducting) state, albeit
with only quasi long-range order within the Beresinskii-Kosterlitz-Thouless (BKT) picture~\cite{Berezinskii72,Kosterlitz73}. 
 In Fig.~\ref{fig:one} we also included,
as a dashed line, the mean-field transition temperature ({\em not\/} the BKT transition temperature) on the same plot, 
which satisfies 
\be
\label{eq:mftc}
 0 = \int_0^\infty d\epsilon \Big( \frac{\tanh \frac{1}{2}\beta (\epsilon - \mu)}{2(\epsilon-\mu)} 
 - \frac{1}{2\epsilon + \epsilon_b} \Big) .
\ee
Within mean-field theory, pairing correlations are expected to become strong to the right of this curve, and it is interesting
to note that the peak  in $n/n_0$ for the various curves approximately coincides with the mean-field transition.  Physically, this 
indicates that interaction effects are strongest for temperatures just above the mean-field transition temperature, although we must
again note that the true BKT transition temperature at which superfluidity is formed would occur at even lower temperatures, further
to the right in Fig.~\ref{fig:one}.

\begin{figure}[ht!]
     \begin{center}
        \subfigure{
            \label{fig:3a}
            \includegraphics[width=85mm]{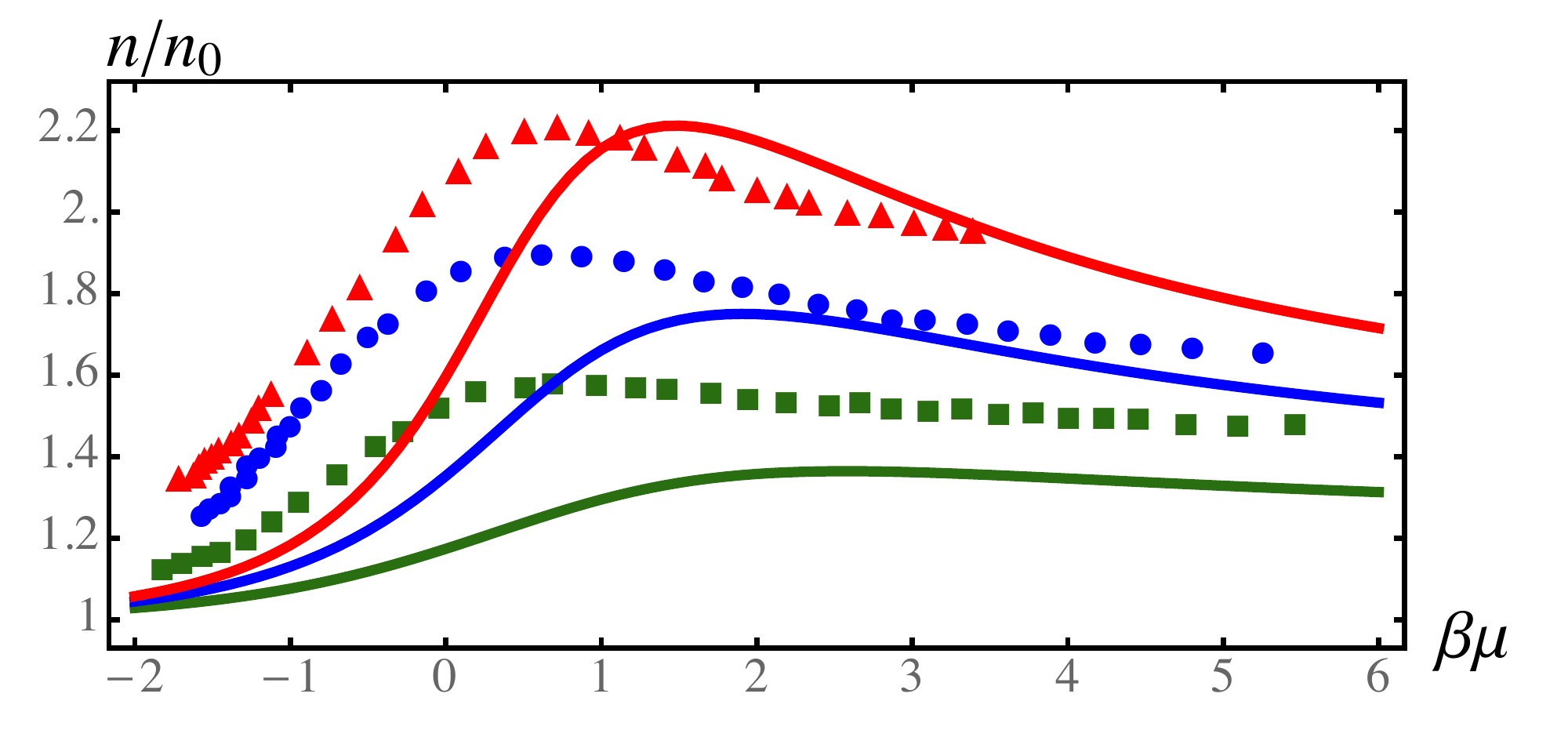}}
        \\ \vspace{-.5cm}
        \subfigure{
            \label{fig:3b}
            \includegraphics[width=85mm]{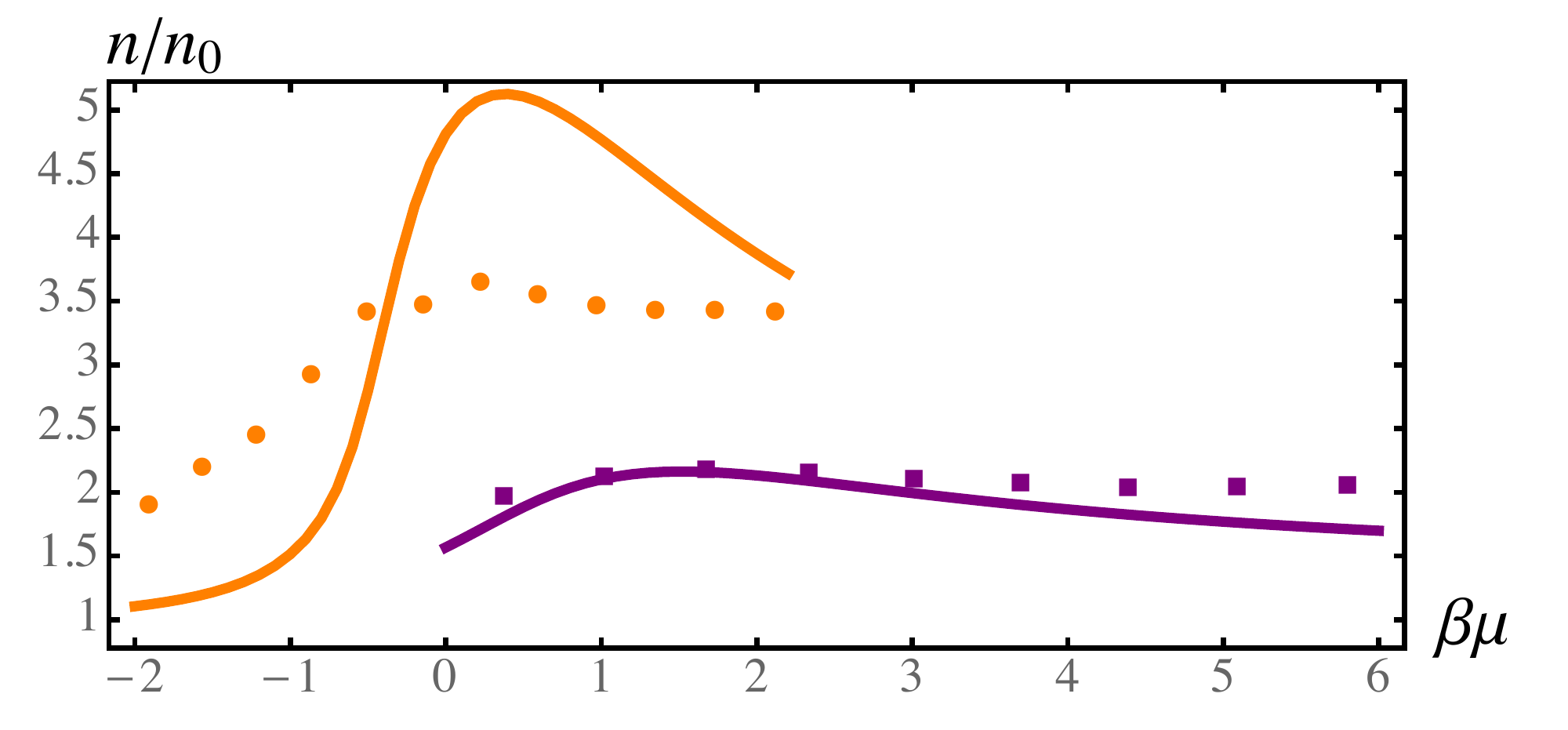}}
    \end{center}\vspace{-.5cm}
    \caption{
      (Color Online) The top panel shows the comparison of  Eq.~(\ref{eq:finalmuvsn})
      (solid curves) with the data of Fenech et al~\cite{Fenech16} 
(points) for coupling values the same as in Fig.~\ref{fig:two}.  The bottom panel compares 
Eq.~(\ref{eq:finalmuvsn}) (solid curves) with
the data of Boettcher et al~\cite{Boettcher16}, with the top (orange) plots being $\beta E_b = 1.2$ and the bottom (purple)
plots being $\beta E_b = 0.5$.}
\vspace{-.5cm}
   \label{fig:three}
\end{figure}

We conclude this section by comparing our theory to the experiments of Refs.~\cite{Fenech16,Boettcher16}, each of which extracted the density
vs. chemical potential for a uniform 2D gas using the LDA on a trapped 2D Fermi gas.  The top panel of Fig.~\ref{fig:three}
  shows our comparison to the Vale group
data~\cite{Fenech16} for coupling values  $\beta E_b = 0.47$ (top, red)
$\beta E_b = 0.26$ (middle, blue), and $\beta E_b = 0.06$ (bottom, green).   The theory agrees reasonably well with this data, with 
the main difference being that the 
theory curves seem be shifted (along the $\beta \mu$ axis) relative to the experimental data.  In the bottom panel of Fig.~\ref{fig:three},
we compare to the results of the Jochim group~\cite{Boettcher16}, whose experiments are, typically, at larger values of the dimensionless
coupling.  While some agreement holds for the bottom curves at $\beta E_b = 0.5$,  the upper curves,
at the stronger coupling value of $\beta E_b = 1.2$, do not show agreement.  This indicates that the validity of our theory is restricted to smaller
values of the dimensionless coupling, which is sensible since it is based on a perturbative RG calculation.

%

\section{Other observables}
\label{OTQ}
\begin{figure}[ht!]
     \begin{center} 
\label{fig:4}
            \includegraphics[width=85mm]{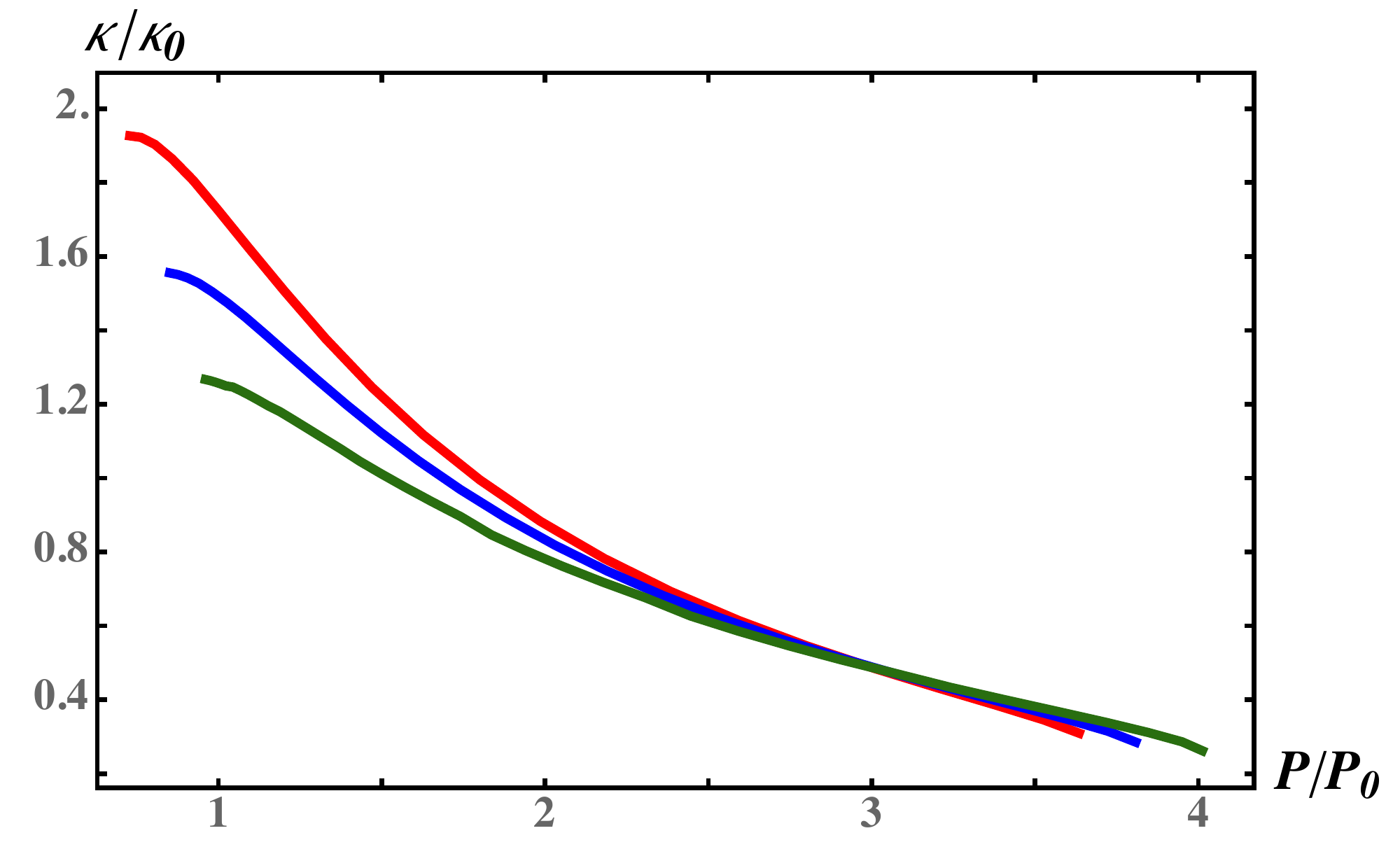}
        \end{center}\vspace{-.5cm}
    \caption{ (Color Online) Normalized isothermal compressibility is
plotted as a function of normalized pressure for interaction strengths
characterized by $\beta\epsilon_b=0.06, 0.26, 0.47$ from bottom to top starting from the left 
 (green, blue and
red dashed lines, respectively).  } \vspace{-.5cm}
   \label{fig:four}
\end{figure}

Having presented our theoretical results for the density vs. chemical
potential for a balanced 2D gas in the normal phase, in this section
we turn to other thermodynamic observables.  As discussed in Ref.~\cite{Fenech16},
these can be extracted using the Gibbs-Duhem relation $Nd\mu = -SdT + AdP$
with $S$ the entropy, $A$ the system  area, $P$ the pressure,
and $N$ the total particle number.  Applying these relations at
constant $T$ allows us to relate
 the density $n=N/A$ to observables like $p$ and the
isothermal compressibility $\kappa = - \frac{\partial n^{-1}}{\partial \mu}$:
\bea
P\lambda_T^4&=&2\pi \int_{-\infty}^{\beta\mu} n(\beta\mu^\prime) \lambda_T^2 d(\beta\mu^\prime),
\\
\kappa&=& \frac{\lambda_T^4}{2\pi}
\frac{1}{(n(\beta\mu)\lambda_T^2)^2} \frac{d(n(\beta\mu)\lambda_T^2)}{d(\beta\mu)},
\eea
where we introduced $\lambda_T=\sqrt{2\pi/(m  T)}$, the thermal de Broglie wavelength,
to write all quantities
in dimensionless forms.  In Fig.~\ref{fig:four}, we plot our results for the
compressibility vs. pressure for
three values of the dimensionless interaction, each normalized to their non-interacting counterparts,
an observable that is expected to show signatures of the phase transition into the superfluid
state~\cite{Ku2012}
\bea
\kappa_0&=&\frac{1}{\lambda_T^2} \frac{e^{\beta\mu}}{e^{\beta\mu}+1}, \\
P_0&=& -\frac{2}{\beta \lambda_T^2} {\rm Li}_2 (-e^{\beta\mu}+1),
\eea
with ${\rm Li}_n(z)$ being the polylogarithm function.  Although we have not attempted a detailed comparison,
the behavior of Fig~\ref{fig:four} is qualitatively consistent with the data of Ref.~\cite{Fenech16}.

\section{Concluding Remarks}
\label{sec:CR}
In this paper we have used a renormalization group (RG) approach to understand
recent experiments that measured the density equation of state for
interacting fermionic atomic gases, observing a bump in the normalized
density $n/n_0$ as a function of chemical potential, with $n_0$ the
noninteracting density~\cite{Fenech16, Boettcher16}.  We presented a simplified
RG calculation based on terminating the coupling constant
RG equation at a density or temperature
dependent scale, which qualitatively reproduces the observed bump.   A more
detailed analysis based on integrating the full RG equation yielded
a result with a momentum-dependent effective coupling in the equation of
state.  This analysis showed reasonably good agreement with experimental
data.  Our resulting picture is that the bump represents a logarithmic correction to the noninteracting
equation of state (expected for a system with a marginal coupling), with 
the effective renormalized coupling crossing over from $\lambda_{\rm eff}~\sim 1/\ln T$ 
at $\frac{\pi n}{mT}\ll 1$ in the temperature-dominated regime to $\lambda_{\rm eff}~\sim 1/\ln n$
at $\frac{\pi n}{mT} \gg 1$ in the density-dominated regime.  

We note that other theoretical approaches have obtained the
observed bump~\cite{Fenech16, Boettcher16}, including Ref.~\cite{Bauer14}
using a self-consistent $t$ matrix approach and Ref.~\cite{Anderson15} using lattice Monte Carlo.
Our approach has the advantage of being simpler than these heavily numerical
approaches while also providing reasonable agreement with experiments.  Future
possible research directions include extending this approach to the superfluid
phase and to the imbalanced regime of 2D Fermi gass.

We gratefully acknowledge useful discussions and correspondence with
the research groups of C. Vale and S. Jochim who generously provided
data.  We also acknowledge useful discussions with J. Schmalian.
This work was supported by the National Science Foundation Grant No.
DMR-1151717.  Part of  this work was performed at the Aspen Center for
Physics, which is supported by National Science Foundation grant
PHY-1607611.


\begin{thebibliography}{10}

\bibitem{Bloch08} 
I. Bloch, J. Dalibard, and W. Zwerger,
Rev. Mod. Phys. {\bf 80}, 885   (2008). 
\bibitem{Giorgini08} 
  S. Giorgini, L. P. Pitaevskii, and S. Stringari,  Rev. Mod. Phys. \textbf{80}, 1215 (2008).
  
\bibitem{LevinsenReview15} J. Levinsen and M. M. Parish, 
Annual Review of Cold Atoms and Molecules {\bf 3}, 1 (2015).

\bibitem{Modugno03} 
G. Modugno, F. Ferlaino, R. Heidemann, G. Roati and M. Inguscio, Phys. Rev. A. {\bf 68}, 011601 (2003).

\bibitem{Feld11}
M. Feld, B. Fr\"ohlich, E. Vogt, M. Koschorreck, and M. K\"ohl, Nature {\bf 480}, 75 (2011).

\bibitem{Sommer12} 
A.T. Sommer, L.W. Cheuk, M.J.H. Ku, W.S. Bakr, and M.W. Zwierlein,
Phys. Rev. Lett. {\bf 108}, 045302 (2012).

\bibitem{Zhang12}
Y. Zhang, W. Ong, I. Arakelyan, and J.E. Thomas, Phys. Rev. Lett. {\bf 108\/}, 235302 (2012).
%

\bibitem{Vogt12} 
E. Vogt, M. Feld, B. Fr\"ohlich, D. Pertot, M. Koschorreck, and M. K\"ohl, Phys. Rev. Lett {\bf 108}, 070404 (2012).

\bibitem{Makhalov2014} 
V. Makhalov, K. Martiyanov, and A. Turlapov, Phys. Rev. Lett. {\bf 112\/}, 045301 (2014). 
%
%
\bibitem{Ries15}
M.G. Ries, A.N. Wenz, G. Z\"urn, L. Bayha, I. Boettcher, D. Kedar, P.A. Murthy, M. Neidig, T. Lompe, and S. Jochim, 
Phys. Rev. Lett. {\bf 114}, 230401 (2015).
\bibitem{Berezinskii72} V.L. Berezinskii, 
%
%
Sov. Phys. JETP {\bf 34}, 610-616 (1972).
\bibitem{Kosterlitz73}
J. M. Kosterlitz and D. J. Thouless, J. Phys. C: Solid State Phys. {\bf 6}, 1181 (1973).

\bibitem{Murthy15}
P. A. Murthy, I. Boettcher, L. Bayha, M. Holzmann, D. Kedar, M. Neidig, M. G. Ries, A. N. Wenz, G. Z\"urn, and S. Jochim,
Phys. Rev. Lett. {\bf 115}, 010401 (2015).
\bibitem{Cheng2016} C. Cheng, J. Kangara, I. Arakelyan, and J.E. Thomas, Phys. Rev. A {\bf 94}, 031606(R) (2016).
\bibitem{Mitra2016} D. Mitra, P.T. Brown, P. Schau\ss, S.S. Kondov, and W.S. Bakr,
  Phys. Rev. Lett. {\bf 117}, 093601 (2016). 
\bibitem{Holten2018} 
M. Holten, L. Bayha, A. C. Klein, P. A. Murthy, P. M. Preiss, and S. Jochim,
Phys. Rev. Lett. {\bf  121\/}, 120401 (2018)
\bibitem{Peppler2018}
T. Peppler, P. Dyke, M. Zamorano, I. Herrera, S. Hoinka, and C.J. Vale, 
Phys. Rev. Lett. {\bf  121\/}, 120402 (2018).
\bibitem{Gaunt} 
A. L. Gaunt, T. F. Schmidutz, I. Gotlibovych, R. P. Smith, and Z. Hadzibabic, 
Phys. Rev. Lett. {\bf 110}, 200406 (2013).
\bibitem{Mukherjee17} 
B. Mukherjee, Z. Yan, P.B. Patel, Z. Hadzibabic, T. Yefsah, J. Struck and M. W. Zweirlein,
Phys. Rev. Lett {\bf 118}, 123401 (2017).
\bibitem{Hueck18} 
K. Hueck, N. Luick, L. Sobirey, J. Siegl, T. Lompe and H. Moritz, Phys. Rev. Lett {\bf 120}, 060402 (2018).
%
\bibitem{Randeria89}
  M. Randeria, J.-M. Duan, and L.-Y. Shieh, Phys. Rev. Lett. {\bf 62}, 981 (1989).
  %
\bibitem{Randeria90} 
M. Randeria, J. Duan and L. Shieh, Phys. Rev. B {\bf 41}, 327 (1990).
\bibitem{Petrov03}
D.S. Petrov, M.A. Baranov, and G.V. Shlyapnikov, Phys. Rev. A {\bf 67}, 031601(R) (2003).
\bibitem{Fischer13}  
A. M. Fischer and M. M. Parish, Phys. Rev. A {\bf 88}, 023612 (2013).
\bibitem{Fenech16} 
K. Fenech, P. Dyke, T. Peppler, M.G. Lingham, S. Hoinka, H. Hu, and C.J. Vale, Phys. Rev. Lett. {\bf 116}, 045302 (2016).
%
\bibitem{Boettcher16} 
I. Boettcher, L. Bayha, D. Kedar, P. A. Murthy, M. Neidig, M. G. Ries, A. N. Wenz, G. Z\"urn, S. Jochim, and T. Enss, 
Phys. Rev. Lett. {\bf 116}, 045303 (2016).
\bibitem{Sheehy2007} D. E. Sheehy and J. Schmalian,
Phys. Rev. Lett.
\textbf{99}, 226803 (2007).

\bibitem{Bloom74} P. Bloom, Phys. Rev. B { \bf 12}, 125 (1975).
%
%

\bibitem{Ku2012}
M. J. H. Ku, A. Sommer, L. W. Cheuk, and M. W. Zweirlein, Science {\bf 335}, 563 (2012).
%
%
%
\bibitem{Luttinger}
J.M. Luttinger and J. C. Ward,
 Phys. Rev. {\bf 118}, 1417 (1960).

\bibitem{Bauer14} 
M. Bauer, M. M. Parish and T. Enss, Phys. Rev. Lett {\bf 112}, 135302 (2014).
%
\bibitem{Anderson15}
  E. R. Anderson, and J. E. Drut, Phys. Rev. Lett. {\bf 115}, 115301 (2015).
%

\end{thebibliography}
\end{document}